\documentclass[manuscript]{acmart}
\usepackage{graphicx}
\usepackage{subcaption}
\AtBeginDocument{%
  }

\newcommand{\system}{\textsc{ORE}}
\begin{document}

\title{\textsc{Open Response Explorer}: a Structured Collaborative Writing System for Peer-Informed Open-Response Assignments}

\author{Ziwei Gu}
\email{ziweigu@g.harvard.edu}
\affiliation{%
  \institution{Harvard University}
  \city{Cambridge}
  \state{Massachusetts}
  \country{USA}
}

\author{Ruixuan Chen}
\email{ruixuan68@gmail.com}
\affiliation{%
  \institution{Santa Monica College}
  \city{Santa Monica}
  \state{CA}
  \country{USA}
}

\author{Elena Glassman}
\email{glassman@seas.harvard.edu}
\affiliation{%
  \institution{Harvard University}
  \city{Cambridge}
  \state{Massachusetts}
  \country{USA}
}


\newif\ifshownotes
\shownotestrue  

\ifshownotes
  \newcommand{\zgnote}[1]{\textcolor{Purple}{[ZG: #1]}}
  \newcommand{\egnote}[1]{\textcolor{pink}{[EG: #1]}}
\else
  \newcommand{\zgnote}[1]{}
  \newcommand{\egnote}[1]{}
\fi

\begin{abstract}
Peer-informed open-response assignments---where students view, connect, and learn from peer contributions---hold the potential to foster collaborative knowledge building, yet the unstructured nature of free-form prose prevents students from effectively comparing and making sense of peers' work. We present \textsc{Open Response Explorer}, an interface that recenters open responses on structured idea units through purpose-based writing and peer response exploration, augmented by common-role text segment highlighting for easy comparison. A within-subjects lab study (N=12) validated the usefulness and usability of the system, demonstrating how it shaped the open-response drafting and peer exploration practices. We discuss implications for purpose-centered authoring and exploration that transform free-form prose into structured, comparable, and collectively navigable knowledge artifacts.

\end{abstract}

\maketitle

\section{Introduction}
\label{sec:intro}

Open-response assignments---such as paper reviews, case briefs, design justifications, and short-answer explanations---are central to learning across K--12, undergraduate, and graduate education. They create space for explanation, argumentation, and reflection, and are traditionally performed privately to assess individual mastery. However, evolving classroom designs~\cite{chan2018knowledge} have also explored the use of \textit{public} open responses for collaborative knowledge building in communities of learners~\cite{scardamalia2006knowledge, tarchi2013knowledge}. For example, discussion forum posts invite learners to contribute ideas in their own words, encounter diverse perspectives, and refine their ideas through social exchange~\cite{onyema2019online}, a practice shown to improve course performance~\cite{cheng2011assessing}. Decades of research in social learning and studio pedagogy\footnote{An educational model where learning occurs through the public construction of artifacts and visible peer critique rather than private study~\cite{schon1987educating}} suggest that this visibility allows students to view, compare, and model their work after the diverse strategies of their peers, rather than working in isolation~\cite{bandura1977social, schon1987educating}, thereby transforming a classroom of individuals into a community of inquiry~\cite{onah2014exploring}. Drawing on this theoretical grounding, we envision a hybrid model we term \textit{peer-informed open-response assignments}---tasks where students respond to a shared prompt but are encouraged to view peer responses at some stage of the assignment lifecycle (before, during, or after drafting) to leverage the public corpus for \emph{comparative reflection}. 

However, realizing the social and pedagogical potential of this hybrid model is challenging due to the \emph{unstructured messiness} of raw prose. Existing peer learning platforms (e.g., SWoRD~\cite{cho2007sword}, Calibrated Peer Review~\cite{robinson2001calibrated}, PeerStudio~\cite{kulkarni2015peerstudio}, and Perusall~\cite{perusall_site}) scaffold feedback workflows and can improve revision and performance, but lack the affordances to make large volumes of text navigable or comparable because they typically operate at the \emph{document} level and cannot decompose long text into granular units of meaning to allow for direct, like-with-like comparison across peers. Automated essay evaluation can assign scores or surface features of writing at scale~\cite{shermis2013handbook}, but scoring does not, by itself, help students \emph{compare} and \emph{extend} ideas across peers---a cognitive activity that promotes learning useful abstractions~\cite{gentner2017}. Argument mining methods can detect claims and evidence~\cite{stab2017parsing}, and Large Language Models (LLMs) can extract themes from peer responses, but these techniques seldom appear in interactive interfaces that help learners explore peer ideas at scale, contextualize their own responses, and compose with comparison in mind. Consequently, when students (and instructors) attempt to navigate peer contributions, hundreds of heterogeneous responses can quickly become cognitively overwhelming, making it difficult to identify similar or contrasting ideas, trace strong and weak arguments, or build on promising directions in others' work. As a result, peer engagement often collapses into ad hoc commenting, while grading and feedback remain extremely difficult and inconsistent~\cite{onyema2019online, andresen2009asynchronous}.

To address these challenges, we present \textsc{Open Response Explorer} (\system{}), a novel interface designed specifically for peer-informed open-response assignments, where students benefit from seeing, comparing, and building on classmates' ideas. \system{} reduces the intrinsic cognitive demands of reading and reasoning about the emerging set of peer contributions by making idea units first-class, comparable objects. \system{} provides (1) \textbf{Structured Authoring} to a shared prompt, where students compose text in linear, purpose-labeled fields rather than a single text blob, yielding responses that are structured by design; (2) \textbf{Structured Exploration with Common-Role Text Segment Highlighting}, an exploratory view that organizes the class corpus by purposes and highlights analogous segments in distinct colors to facilitate side-by-side comparison; and, either directly from that exploratory view or via one-click retrieval of the pre-existing comment that is most similar to a student's on-going draft, (3) \textbf{Peer-Linked Authoring with Common-Role Text Segment Highlighting}, which
allows students to write commentary on a particular peer's response that \emph{contrasts} with or \emph{extends} it with the assistance of that role highlighting. Together, these interactions reframe open-response work from solitary text production to purpose-centered collaborative knowledge construction and exploration.

To evaluate \system{}, we conducted a within-subjects lab study with 12 participants. The results showed that participants found the system to make it easier to write, explore, and learn from a peer-informed open response assignment. Participants particularly appreciated how structured authoring, coupled with common-role text segment highlighting, facilitated writing, comparing, and sensemaking of peer responses at scale.

In summary, we contribute:
\begin{enumerate}
    \item Novel purpose-centered structured authoring and exploration with role-based highlighting and peer-linked authoring to support peer-informed open-response assignments; 
    \item \system{}, which instantiates these techniques for writing and exploring open-ended responses; 
    \item A within-subjects lab study with 12 participants that validated the usability and usefulness of \system{} for a peer-informed open-response assignment.
\end{enumerate}

\section{Related Work}

\subsection{Peer-Informed Open-Response Assignments: Practices, Challenges, and Existing Tools}
Peer-Informed open-response assignments such as discussion forum posts, mock program committee (PC) reviews, collaborative case studies, and in-class formative writing are increasingly common across K–12 and higher education because they elicit personal articulation and, unlike solitary assessments, invite students to view and engage with one another's work, creating opportunities for social learning~\cite{bandura1977social}. The pedagogical logic behind these  ``compose--contextualize--compare--revise'' assignment workflows draws directly from \emph{studio pedagogy}---a model grounded in social learning theory where work is public, and learning occurs through the visible construction of artifacts and peer engagement~\cite{schon1987educating, bandura1977social}. In physical studios (e.g., architecture, art), seeing peer work is not considered cheating, but rather an essential scaffold for students to acquire professional norms and diverse strategies~\cite{kuhn2002software}. 

However, when operationalizing the benefits of social learning within written assignments, effectively translating this ``visible work'' culture to asynchronous text-based environments like discussion boards remains a significant challenge. Unlike visual artifacts that support rapid scanning and comparison, raw prose is linear and opaque; without structural support, the ``studio'' experience often becomes cognitively overwhelming rather than scaffolding. In particular, as the number of responses grows, the free-form nature of prose exacerbates this cognitive bottleneck: peers find it hard to locate comparable ideas in one another’s work~\cite{robinson2001calibrated}, and instructors struggle with grading consistently and providing feedback~\cite{fertalj2022systematic}, often relying on peer or automated grading~\cite{fertalj2022systematic,baral2021improving} and feedback~\cite{kulkarni2015peerstudio,botelho2023leveraging}.

Mainstream course platforms provide basic workflows for peer learning, e.g., peer review, discussions, and inline annotations, but generally leave responses as unstructured text streams. 
For example, Canvas supports peer-review assignments and graded discussions \cite{canvas_peer_review,canvas_discussions}, Blackboard offers threaded discussions \cite{blackboard_discussions,blackboard_app}, and Moodle includes discussion forums and PDF annotation for assignment feedback \cite{moodle_forum,moodle_annotate_pdf}. Social annotation platforms such as Perusall provide helpful structure to student discussions by requiring every comment to be anchored to some point in the assigned reading, or be a response to another comment that is itself anchored to the text. The spatially distinct anchoring of different conversations in the margins and the specificity of the text they are anchored on help students find and append commentary that is relevant~\cite{perusall_site,perusall_review,perusall_psu,perusall_cti}. Recently, LLM-powered systems have also been proposed. For example, \textsc{GLITTER}~\cite{peng2025glitter} supports material-grounded, asynchronous discussion via concept clustering and reflective summaries, but it does not support students in structuring their responses or provide features for sensemaking across similarly-themed responses. Thus, if the responses are complex, nuanced, long, and/or quite numerous, the challenge still remains.

Collectively, these tools offer improvements in participation and feedback pipelines, but fall short of addressing the core challenge of open responses: the inherent complexity and messiness of unstructured text, which is difficult to digest, compare, or synthesize at scale without support.

\subsection{Finding or Creating Structure in Text for Sensemaking at Scale}
A parallel thread in HCI develops interfaces that surface structure across many texts to support comparison, triage, and sensemaking. 
A number of such systems either (1) identify and reify pre-defined shared structure or (2) attempt to help users see structure emerge that no one necessarily explicitly knew in advance. Informed by theories of human concept learning like Variation Theory~\cite{marton2014necessary}, 
\textsc{Mesotext}~\cite{gero2024supporting} introduces text clustering and rendering techniques such as Positional Diction Clustering with `grid' and `interleaved' views of the results to reveal commonalities and variations across many LLM responses. Informed by Structure-Mapping Theory~\cite{gentner2017}, \textsc{AbstractExplorer}~\cite{abstractexplorer} enables comparative close reading across many similarly structured scientific abstracts by combining role highlighting with structure-aligned ordering and alignment. \textsc{CorpusStudio}~\cite{dang2025corpusstudio} leverages the same theories to guide the design of system features that surface emergent writing patterns across many papers in a corpus to help authors outline and write with community norms in mind.
Similar strategies were used in education contexts at scale as well, primarily intended for teachers to gain better awareness of---and give more, better feedback on---the design choices made by their students at much larger scales, e.g., thousands of student solutions in a Massive Open Online Course like in \textsc{OverCode}~\cite{overcode} and \textsc{Foobaz}~\cite{foobaz}.

Analogous ideas appear in code domains as well: \textsc{ParaLib}~\cite{yan2022concept} uses the same theories to justify concept annotations and color-consistent highlighting to help users mentally align semantically analogous code segments across libraries, and \textsc{Examplore}~\cite{glassman2018visualizing} visualizes API usage examples in Java at scale by displaying their (canonicalized) variation within one shared structure---a fixed code skeleton based on the authors' knowledge of the possible code structures around API calls in Java. Structured code editors like \textsc{Javardise}~\cite{10.1145/3397537.3397561} also uses structure to help beginners learn programming by constraining editing to syntactically valid program code. 

Designers of co-adaptive machine teaching have engaged with these techniques as well. In qualitative analysis, which is conceptually closer to the assessment side of open response assignments, \textsc{PaTAT}~\cite{gebreegziabher2023patat} synthesizes human-interpretable textual patterns to support human-AI collaborative coding (i.e., assigning descriptive labels to individual textual items in a collection) and codebook evolution. Codebook evolution is a form of human-AI co-adaptive machine teaching, and there too the strategy of \textit{generating structured text} has been found to be helpful to those reviewing and assessing them, i.e., in \textsc{MOCHA}~\cite{gebreegziabher2025supporting}---where the strategy was designed to clarify the human's mental model of the code's definition. In our work, rather than an AI generating structured text, the students doing the collective sensemaking are. And while prior learnersourcing work like~\citet{glassman2016learnersourcing} enforced a structure to help students provide effective hints to each other, our work opens this strategy up to collaborative discussion together for open response environments, not programming problems that are highly constrained by needing to pass particular test cases.

In summary, across these systems, the structuring process typically happens \emph{downstream}---transforming or annotating already-written text or code to expose latent structure. In contrast, our work expands what we know about the \emph{upstream} production of text: by introducing structured authoring where students write in typed \emph{fields} (or ``idea units''), \system{} makes structure explicit at input time so that output is structured by \textit{design}, which can then be leveraged by later exploration, comparison, and feedback processes with less heavy post-processing.

\subsection{Role- and Semantic-based Highlighting} 
Semantic role labeling is a well-studied task in natural language processing that assigns roles (e.g., Agent, Patient, Instrument) to spans in text, underpinning many downstream applications~\cite{pradhan2005semantic,chen2025semantic}. Though approaches have evolved from feature-rich classifiers~\cite{pradhan2005semantic} to neural span-based and transformer models~\cite{tan2018deep,fei2021encoder}, classical pipelines that combined parse-based features with machine learning that established early baselines remain relevant~\cite{he2017deep}.

HCI research has also focused on text highlighting as a visualization technique, which has been shown to direct focus and attention productively according to eye-tracking studies~\cite{chi2007visual}. Another recent study on constrained highlighting also demonstrates that some highlights can improve reading comprehension compared to no highlight~\cite{joshi2024constrained}, suggesting that role-aware highlighting may productively steer attention. In the context of data annotation interfaces, \textsc{Context}~\cite{ara2024closing} compares highlight- versus reasoning-based designs and finds that newcomers benefit from contextual aids layered atop highlights, suggesting that exposing semantic role labels to novice users may be useful.

Numerous prior HCI systems have utilized text highlighting as a means to facilitate sensemaking. Early work on skimming support introduced machine-learned or human-annotated highlights to emphasize salient parts of sentences. For example, \textsc{VarifocalReader}~\cite{koch2014varifocalreader} highlights key sentence segments in different colors to aid rapid reading. Building on this idea for scholarly reading, \textsc{Scim}~\cite{fok2023scim} helps readers skim academic papers by using colored highlights to guide attention to predefined content types such as ``Objective'' and ``Method.'' \textsc{ClioQuery}~\cite{handler2022clioquery} assists historians with exploring queries in context through linked views coupled with text highlighting. Within the broader Semantic Reader project~\cite{lo2023semantic}, \textsc{Relatedly}~\cite{palani2023relatedly} highlights unexplored, dissimilar information in related work sections of unread papers to steer attention to previously unseen content. Finally, \textsc{AbstractExplorer}~\cite{abstractexplorer} segments sentences into grammar-preserving chunks and assigns each chunk a functional role with color highlights, which becomes the basis for structure-finding and text alignment that supports comparative close reading of thousands of scientific paper abstracts.

Our \system{} builds on this trajectory and applies role-based highlighting to structured student-authored open responses, so that the actual text the student wrote can be understood, compared, assessed, and/or responded to more efficiently.

\section{\system{}}
\subsection{Design Goals and Rationales}
\label{sec:design-goals}
To tackle the challenges of making sense of large text corpora that undermine the learning potential of peer-informed open-response assignments, we synthesized findings from prior research on learning science, sensemaking systems, and related human cognitive theories, supplemented by our classroom observations, to derive four design goals for \system{}.

\noindent\textbf{G1. Breakdown monolithic single responses into a sequence of separate purpose-specific fields}
Unstructured open responses are difficult to process for both students and instructors, especially as their number grows. They quickly accumulate into impenetrable piles, since each response must be consciously interpreted on its own. As a result, people typically read them sequentially, relying on memory and any mental schema built from prior responses to interpret each new one~\cite{rayner2012psychology}. This makes it especially challenging to identify relationships---such as common themes or outliers across students' responses---or to situate a particular response within the context of a larger class. 

Prior research in the learning sciences has proposed idea-based learning~\cite{hansen2023idea} and idea-centered knowledge building~\cite{hong2009towards,hong2019integrated}. In parallel, \citet{weintrop2018block} demonstrated that block-based representations can reduce representational complexity for novice programmers, compared with text-based representations. However, this approach has yet to be specifically applied to written assignments, and the possible analogical transfer of this strategy to natural language assignments would likely only be an incomplete mapping~\cite{gentner2017} since natural language has far more semantic ambiguity than a programming language. 

Research on argument mapping and concept mapping also shows that externalizing various components of a piece of writing, such as claims, evidence, and relations, scaffolds higher-order reasoning and improves clarity~\cite{davies2011concept}. Building on this line of research, we propose a shift from writing free-form prose to a sequence of separate units of prose for statements of different, specific purposes, making ~\textit{purpose fields} the primary entity of composition and comparison in open-response assignments.

\noindent\textbf{G2. Co-define a minimal shared vocabulary of purposes.}
Collaboratively generated classification systems, or \emph{folksonomies}, demonstrate that shared vocabularies can emerge from community tagging rather than top-down taxonomies~\cite{xu2008exploring}. From a distributed cognition perspective, such shared external representations function as collective cognitive artifacts that support group understanding~\cite{hutchins1995cognition}. Crowdsourcing research further shows that complex taxonomies can be constructed through small, distributed contributions~\cite{chilton2013cascade}. We thus intend to enable students to co-define a minimal shared vocabulary of purposes, allowing the schema to evolve with classroom practices and reflect the diversity of perspectives present, instead of having the purposes pre-defined by the instructor.

\noindent\textbf{G3. Support scalable response comparison.}
Grouping sentences by shared purpose (G1.) offers an opportunity to further identify and reify common roles of text segments within each sentence to facilitate cross-sentence \emph{comparison}.
According to Structural Mapping Theory, a well-established framework for human concept learning in cognitive science, understanding and sensemaking arise from mapping relationships between alignable structures, particularly in complex tasks~\cite{gentner1983structure, gentner2001structural}. (Comparison as a means for sensemaking is also instantiated in recent systems that improves close reading at scale~\cite{gero2024supporting,abstractexplorer}.) 
We can provide further support to help readers compare large sets of responses by revealing analogous and conceptually comparable roles that are embedded in prose, especially when the text being directly compared has the same purpose (G1).

\noindent\textbf{G4. Embed peer context into drafting when appropriate.}
In certain types of assignments, peer activity can increase engagement and improve revision quality when integrated during the drafting process~\cite{kulkarni2015peerstudio,cho2007sword}. Knowledge-building perspectives emphasize that learning advances when individuals notice, build on, and differentiate peers’ contributions~\cite{scardamalia2006knowledge}. We thus intend to bring timely, relevant peer responses into the writing context to support building-on and perspective-taking in discussion-based assignments without displacing authorial agency.

\noindent\textbf{G5. Stage peer exposure to balance independence and collaboration.}
We designed the system to support configurable visibility to peer work to serve different pedagogical goals. For assessment-oriented tasks, the system supports independent drafting where students write privately. In this mode, the benefits of structured authoring primarily accrue to the instructor, who can more easily navigate and grade the structured corpus. For collaborative learning tasks, the system supports open or staged access to peer context, where students view peer work either immediately or after independently drafting their own work. This flexibility aligns with frameworks for both productive failure and social constructivism~\cite{kapur2016examining,chi2014icap,mazur1999peer,rudd2009calibrated}. While \system{} enables both workflows, the rest of the paper focuses on the open exploration mode to illustrate and evaluate the more general collaborative sensemaking and knowledge building workflow.

\subsection{Key Features}
\label{sec:key-features}

\system{} comprises four complementary features that, together, address the challenges of making sense of free-form text in open-response assignments and realize our design goals (\S\ref{sec:design-goals}): 
\begin{enumerate}
    \item \textbf{Structured Authoring (purpose-based writing)}, which recenters composition on text units with distinct purposes (G1–G2); 
\item \textbf{Structured Exploration with Common-Role Text Segment Highlighting}, which organizes peer work for like-with-like comparison at scale (G1–G3); 
\item \textbf{Peer-Linked Authoring}, which brings peer context into view when drafting one's own response---when appropriate to support building-on and perspective-taking---while preserving authorship (G4).
\item \textbf{Instructor-configurable Modes}, which allow phasing or disabling peer exposure (e.g., turning off Peer-Linked Authoring or releasing \textbf{Explore} only after submission or only to instructors) to balance independence and collaboration (G5).
\end{enumerate}
\begin{figure}
    \centering
    \includegraphics[width=\linewidth]{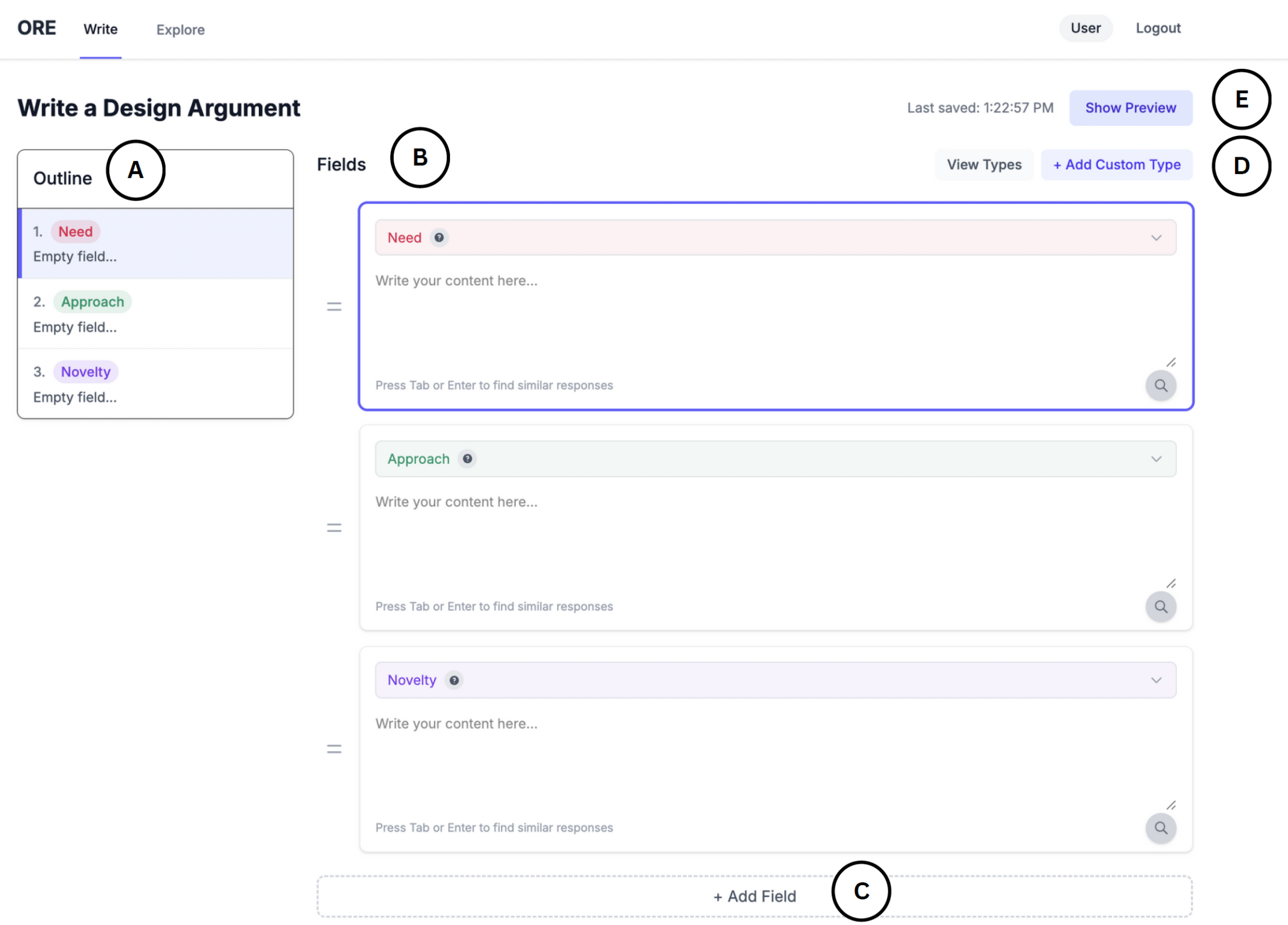}
    \Description[]{}
    \caption{\textbf{Write} view with typed (purpose-specific) fields. (A) Live outline listing each field’s type and opening words; (B) editable field list with color-coded dropdowns of purposes and drag handles; (C) \emph{+ Add Field} appends a new field; (D) \emph{View Types} and \emph{+ Add Custom Type} manage the shared field type palette; (E) \emph{Show Preview} stitches fields into an essay.}
    \label{fig:write}
\end{figure}

\subsubsection{Structured Authoring (purpose-based writing)}
\label{sec:feature-authoring}

To recenter authoring on a sequence of purposes (G1) and enable class-wide collaborative generation of those purposes (G2), \system{} replaces a single text box with a sequence of \emph{purpose fields}. Each field is a distinct unit of thought labeled with a \emph{purpose} (also called its type). The class co-defines a shared palette of field purposes for the assignment, so that purposes are explicit and comparable across students.

Students draft by adding, editing, and reordering fields of various purposes until the response forms a coherent argument. As they compose, the left-hand \emph{outline} (Fig.~\ref{fig:write}\textbf{A}) continuously reflects the evolving structure, listing each field’s purpose and its opening words. The central pane displays the \emph{field list} (Fig.~\ref{fig:write}\textbf{B}). Each field header shows its selected purpose from a dropdown of all available purposes and applies the unique color mapped to that purpose for immediate visual recognition. Fields can be reordered by dragging the handle on the left edge, appended via the \emph{Add field} control at the end of the list (Fig.~\ref{fig:write}\textbf{C}), or removed by clicking the red cross that appears on hover. Type management (Fig.~\ref{fig:write}\textbf{D}) is integrated into the same flow: \emph{View Types} reveals the class-defined palette of purpose options, and \emph{Add Custom Type} lets a student propose a new purpose by providing a descriptive label, after which a new color is assigned automatically. At any point, \emph{Show Preview} (Fig.~\ref{fig:write}\textbf{E}) stitches the fields into a conventional essay that can be read or copied. All changes are autosaved to minimize interruption.

\begin{figure}
    \centering
    \includegraphics[width=\linewidth]{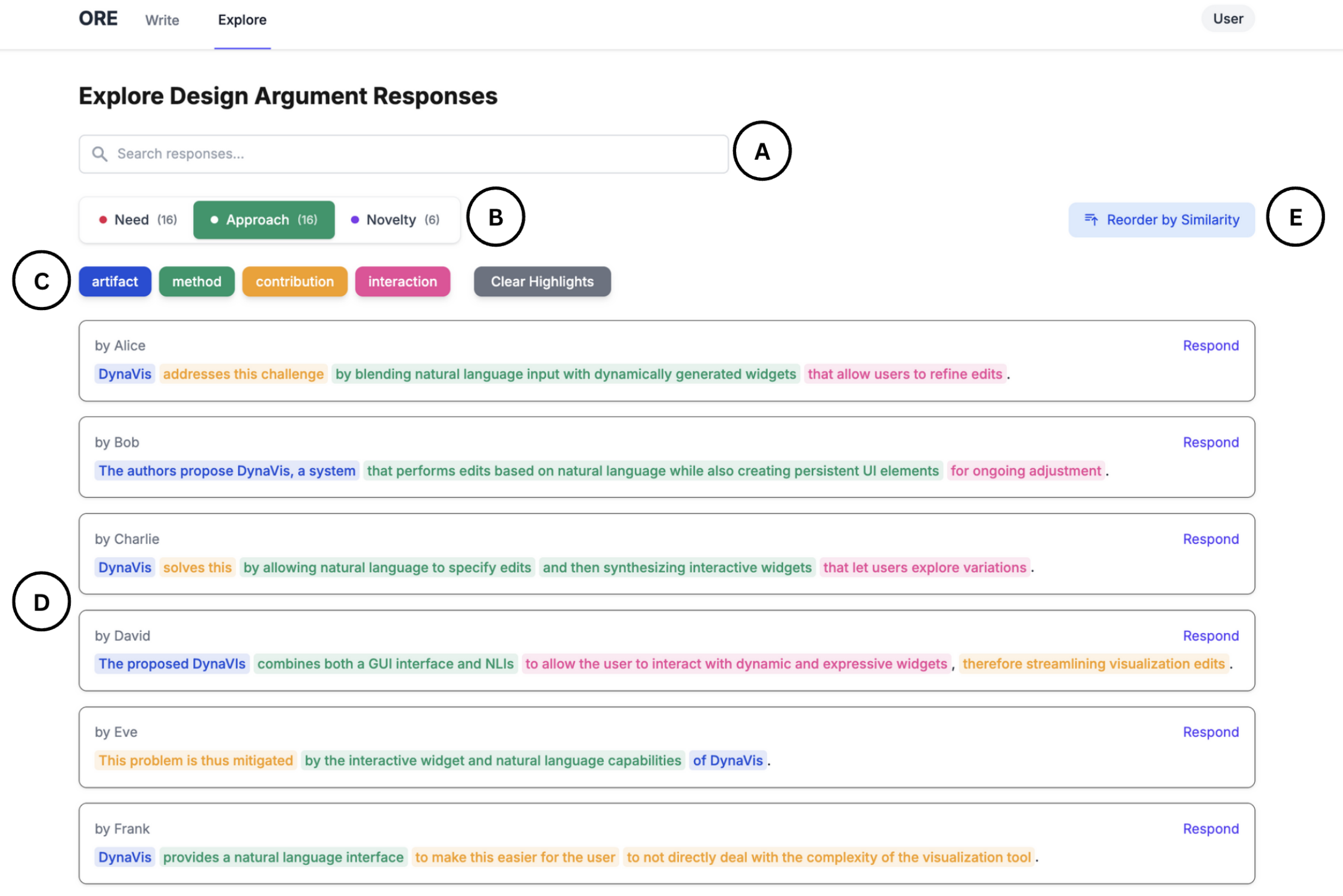}
    \Description[]{}
    \caption{\textbf{Explore} view with common-role text segment highlighting. (A) Keyword search to filter responses; (B) field-type toggles with counts, ordered by prevalence; (C) per-role highlight controls (labels toggle on/off, with \emph{Clear Highlights}; on by default); (D) participants’ responses rendered with consistent role colors and a \emph{Respond} action; (E) reorder responses by similarity to the user’s draft.}

    \label{fig:explore}
\end{figure}

\subsubsection{Structured Exploration with Common-Role Text Segment Highlighting}
\label{sec:feature-exploration}
To enable scalable, like-with-like comparison (G3) across sentences with the same purpose (G1), the \textit{Explore} tab organizes peer work by field type. It also augments each response with \emph{common-role text segment highlights}, informed by Structure-Mapping Theory, which holds that assigning common labels (in this case, assigned by LLMs and indicated with color-coordinated text color and highlighting) to components of an object (i.e., text segments) that play the same role within the structure of that object (i.e., a sentence with a particular purpose),
supports more effective comparison, creating the conditions for learning of more latent, less superficial abstractions~\cite{gentner2017}.

The search bar at the top (Fig.~\ref{fig:explore}\textbf{A}) allows users to filter the corpus by keyword to narrow the exploration space. Users can use the field-type toggle (Fig.~\ref{fig:explore}\textbf{B}) to reveal all responses of a given type, ordered by the number of responses in that type and annotated with counts in parentheses. This way, readers can focus on facets most likely to share structure under Structure-Mapping Theory. A role panel (Fig.~\ref{fig:explore}\textbf{C}) lists all semantic roles that appear at least once within the selected type. Clicking a role toggles its highlighting on or off, and a \emph{Highlight All / Clear Highlights} control switches all highlights together. By default, all highlights are shown. The main response list (Fig.~\ref{fig:explore}\textbf{D}) renders participants’ responses with role-based highlights applied to analogous segments, providing visual cues about what information each response contains (e.g. in argumentation tasks, what stance each writer takes). Each item includes a \emph{Respond} option to start a conversation with the author (more details in \S\ref{sec:feature-peerlinked}). A similarity control button (Fig.~\ref{fig:explore}\textbf{E}) reorders the list by similarity to the user’s own draft to surface the most comparable responses first. If no draft exists yet, the interface presents a gentle warning and retains the current ordering.

\begin{figure}[t]
  \centering
  \begin{subfigure}{0.8\linewidth}
    \includegraphics[width=\linewidth]{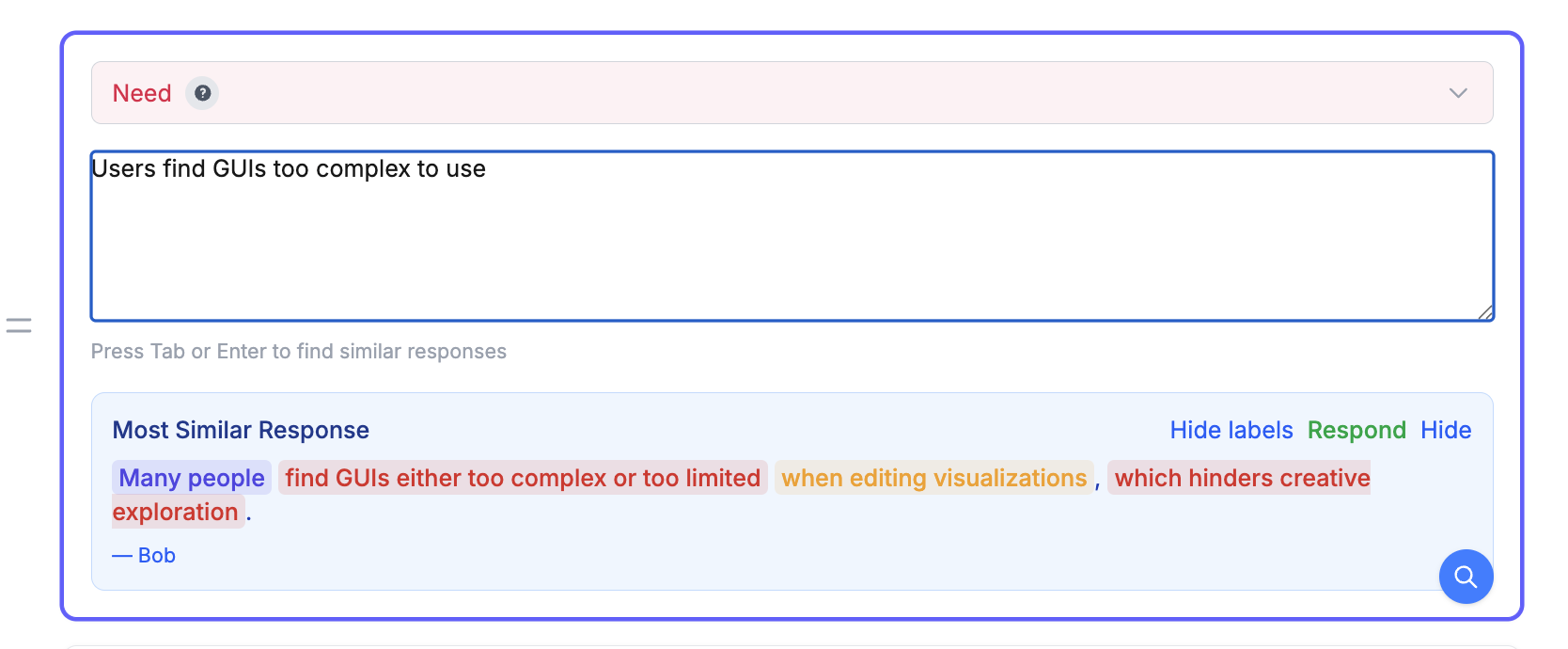}
    \caption{Most-similar peer response surfaced inline.}
    \label{fig:most-similar}
  \end{subfigure}

  \vspace{0.6\baselineskip}

  \begin{subfigure}{0.8\linewidth}
    \includegraphics[width=\linewidth]{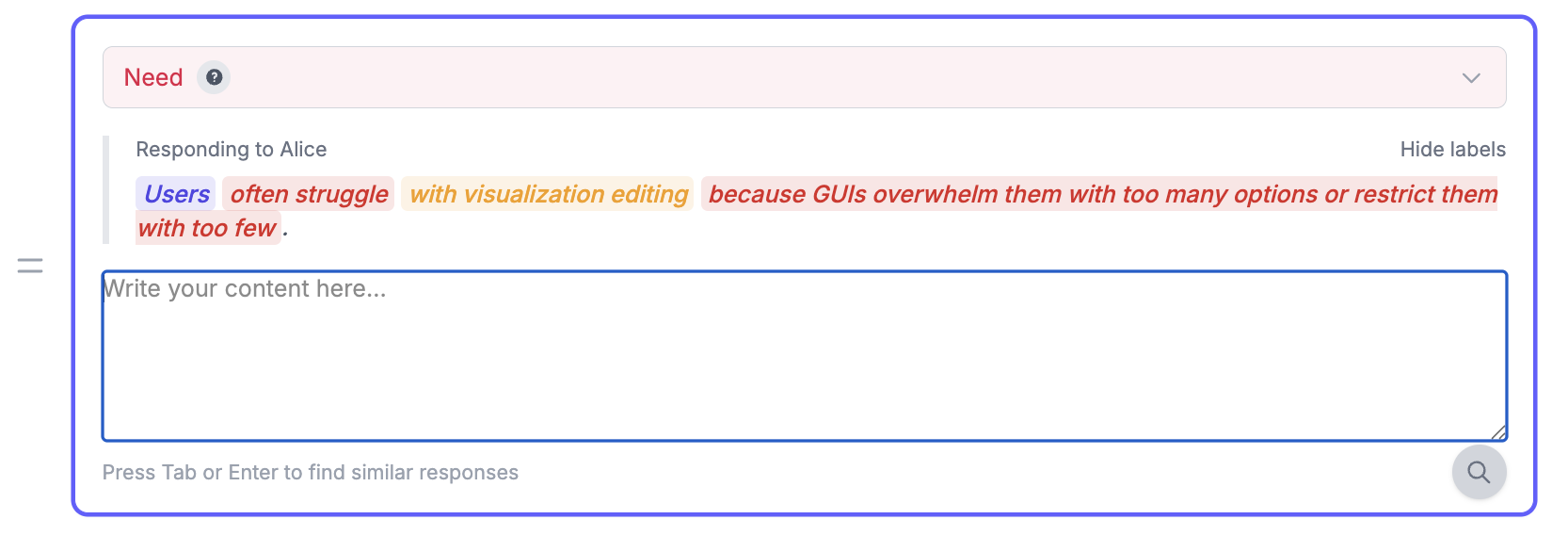}
    \caption{After \emph{Respond}: same-type field created with quoted excerpt.}
    \label{fig:respond}
  \end{subfigure}

  \caption{Peer-Linked Authoring: surfacing the most similar peer response (top) and responding to create a same-type field (bottom).}
  \label{fig:peer-linked}
  \Description{Two screenshots: top shows a writing field with an inline panel containing the most similar peer response; bottom shows the editor after clicking Respond, with a new same-type field added and attributed.}
\end{figure}

\subsubsection{Peer-Linked Authoring}
\label{sec:feature-peerlinked}

To embed timely peer context into drafting (G4), the \emph{Write} interface allows students to surface the most similar peer response to the field they are currently composing. While the cursor is in a field, pressing \emph{Tab} or \emph{Enter} or clicking the \emph{Find} icon reveals a compact view of the nearest peer response alongside the field, together with the author’s name. The response carries over the common-role text segment highlights introduced in \S\ref{sec:feature-exploration} so that the roles of components are clear and consistent with the \emph{Explore} interface (Fig.~\ref{fig:most-similar}). Hovering reveals the corresponding labels, and a \emph{Hide Labels} control turns coloring and labels off if desired. This lightweight affordance provides a low-overhead way to situate one’s writing within the broader class corpus, even when the action is as simple as viewing a single most similar response. If no similar response is available because the field is empty or contains too little text, the interface prompts the student to add more content to the field.

As an alternative entry point, students can \emph{Respond} to any peer response on the \emph{Explore} page. Doing so creates a new field at the bottom of the \textbf{Write} page with the same field type as the selected peer response. The peer’s excerpt is quoted and attributed by name, and the highlight state carries over and remains toggleable (Fig.~\ref{fig:respond}). Across both entry points, Peer-Linked Authoring is intended to facilitate building upon and differentiating peers’ responses with minimal friction while preserving student authorship.

\subsubsection{Instructor-configurable modes.}
To accommodate assignments with different expectations for independence, \system{} provides per-assignment controls. Peer-Linked Authoring (\S\ref{sec:feature-peerlinked}) can be disabled to avoid in-draft peer influence. Access to Structured Exploration with Common-Role Text Segment Highlighting (\S\ref{sec:feature-exploration}) on the \textbf{Explore} page can be scheduled: instructors may enable it during drafting for collective knowledge-building tasks, restrict visibility until after submission for assessment-oriented tasks, or disable it entirely for private assignments. These controls modulate whether and when peer context appears without altering the underlying authoring model.

Together, the \textbf{Write} and \textbf{Explore} views support a compose--contextualize--compare--revise workflow common in peer-informed assignments that aligns with our design goals outlined in \S\ref{sec:design-goals}. On \textbf{Write}, students externalize thoughts as purpose-labeled units, making structure explicit (G1) while co-defining a minimal shared vocabulary for reliable comparison (G2). On \textbf{Explore}, students traverse a purpose-organized peer response corpus with common-role text segment highlights to perform comparisons across (potentially large numbers of) peer responses (G3). Peer-linked actions bridge the two views, bringing relevant peer context into drafting without displacing authorship (G4). Finally, by deferring or disabling \textbf{Explore} access, instructors can enforce independent drafting to prioritize individual assessment or maintain submission privacy. In concert, \system{} transform open-response assignments from isolated prose production into purpose-centered authoring and exploration that we hope better supports the educational goals of diverse types of peer-informed open response assignments.

\subsection{Example Scenario}
\label{sec:scenario}

In a graduate human–computer interaction course, Maya is using \system{} to complete a \emph{Mock PC} assignment. The goal is to write a review of a research paper and, after submitting, read classmates’ reviews to collectively reach an accept/reject decision. Because the instructor requires independent reviewing first, only the Write tab is visible at the beginning. Maya sees that reviews are composed as fields, which prompts her to structure her thoughts. From the field-type dropdown, she finds types already defined by earlier reviewers: ``One-sentence Summary,'' ``Significance of the Problem,'' ``Related Work,'' ``Approach,'' ``Evaluation,'' ``Discussion,'' and ``Suggested Next Steps.'' She begins by adding a ``One-sentence Summary'' and a ``Significance of the Problem'' field, then provides comments on ``Related Work'' and ``Approach.'' When she wishes to comment on writing quality but finds no suitable type, she creates a new type, ``Writing and Style,'' and places her remarks there. This new type then becomes available to the whole class. After writing, she drags fields to refine flow, deletes a redundant field, taps \emph{Show Preview} to read the stitched review, and submits her review.

Upon submission, the Explore tab appears. Twenty-five peers have posted, so there is substantial material to navigate. Curious about what her peers think about the evaluation part of the paper, Maya toggles the ``Evaluation'' type to view all comments about the evaluation study. Automatic role labels for this aspect, including \emph{participant}, \emph{method}, \emph{measure}, \emph{strengths}, and \emph{weaknesses}, highlight analogous segments across reviews, allowing her to quickly scan positive and negative points for ``Evaluation.'' She briefly hides a label to reduce visual load, focusing on \emph{weaknesses} and quickly finds a common pattern. Many reviewers, like Maya, note limited participant diversity and recommend broader age coverage. Spotting a methodological limitation she had not considered, she clicks \emph{Respond} to register agreement and add a short note. She repeats this process for ``Related Work,'' ``Approach,'' and ``Discussion.'' By reordering reviews by similarity to her own, she is able to quickly gauge whether her views are common or outlying. After exploring the class corpus type-by-type, she feels confident about the distribution of opinions and makes an informed accept/reject decision for the Mock PC.

\subsection{Implementation}
\label{sec:implementation}

\system{} is a web application\footnote{The
web application is publicly available at [REMOVED FOR REVIEW]. Our code is publicly available at [REMOVED FOR
REVIEW].} implemented with Next.js (App Router), React, and TypeScript, styled with Tailwind CSS. We use Zustand, a lightweight client storage, for local editor states. Drafts autosave locally via a debounced POST to the API for reload during the session. Usernames, field types, and submitted responses persist in a MongoDB database for continued access.

In the Explore page, common-role text segment highlights are derived using a light-weight, human-in-the-loop pipeline. First, a phrase splitter segments each sentence into clauses and base phrases including noun/verb chunks, producing candidate spans. Next, a Large Language Model (OpenAI GPT-5) assigns a provisional role to each span using few-shot prompts scoped to each field type. To avoid label proliferation, we normalize synonymous labels via a small merge list and collapse adjacent spans with identical labels. A rater then reviews the merged label set for the assignment and approves the final labels. We then run a second pass in which GPT-5 performs multi-class classification over the approved label set, assigning exactly one role per span. Conflicts such as overlapping spans or low-confidence assignments are resolved by preferring longer spans and deferring uncertain cases to the unlabeled state. The resulting spans are rendered with label-consistent colors across the Write and Explore pages. Similarity is computed on the client side as cosine similarity over OpenAI's latest text embeddings (text-embedding-3-large).

\section{User Study}
\label{sec:user-study}
To evaluate the usability, usefulness, and effectiveness of \system{}, we conducted a lab user study\footnote{The study protocol was reviewed and approved by the IRB at the first author's institution.} with 12 participants.  We are mainly interested in answering the following three research questions: 

\begin{itemize}
  \item \textbf{RQ1:} How does \system{} influence how students write, explore, and learn from open-response assignments?
  \item \textbf{RQ2:} What is the perceived usability and utility of \system{}’s features?
  \item \textbf{RQ3:} How effective is \system{} at supporting students' learning and feedback giving in open-response assignments?
\end{itemize}

\subsection{Participants}

We recruited 12 participants (5 male, 7 female) from an R1 university through departmental mailing lists. 
Seven of the 12 participants were between the ages of 25 and 34, while the remaining five were younger (18--24). The majority (six) were PhD students, five were Master's students, and one was a Post-Doc researcher. Participants reported using a variety of tools to complete open-response assignments, including (from most frequent to least frequent) Word processors (e.g., Microsoft Word, Google Docs), collaborative writing platforms (e.g., Notion, Overleaf), AI writing/feedback tools (e.g., ChatGPT, Claude), learning management systems (e.g., Canvas, Blackboard, Moodle), grading/annotation tools (e.g., Gradescope, Turnitin), online survey or form tools (e.g., Google Forms, Qualtrics), handwritten responses (paper, tablet stylus), and collaborative annotation tools (e.g., Perusall).

\subsection{Study Setup}
\subsubsection{Task}

Participants completed two short open-ended writing tasks, one on each interface condition. They were each given a task description script (see Appendix \ref{sec:task}). For each task, participants were told to imagine a course assignment in which they write a \emph{design argument}\footnote{A design argument is a short essay that makes the case for a proposed tool or approach by articulating the need for the tool and how the proposed approach uniquely addresses the need in context.}~\cite{rees2023encouraging} in response to a prior HCI research paper, which is a typical peer-informed open-response assignment scenario that surfaces core affordances of \system{}. To simulate a peer-informed learning environment (e.g., a visible discussion board), several classmates' responses were made available as references that may or may not fully model good practice. A brief primer on design arguments was provided to each participant as well. 

For each task, participants were instructed to: (1) read the Abstract (and optionally the Introduction section) of the research paper, (2) review their classmates’ responses, and (3) write the best short design argument (2-5 sentences) they can. Since participants were not expected to have prior familiarity with design arguments, this workflow was chosen to mimic a scaffolding scenario where learners leverage peer models to overcome the cold start problem and internalize the genre norms of the assignment before attempting their own contribution.

Each task lasted 15 minutes total, including both exploration of classmates’ responses and writing. Participants were allowed to draw language from the assigned paper when helpful (copy–paste phrases). The task instructions also emphasized that (1) there are no right or wrong answers; (2) the study evaluates the interfaces' effectiveness, not participants’ intelligence; and (3) participants may agree or disagree with the paper and/or classmates’ responses.

\subsubsection{Data}
To simulate a populated learning environment and overcome the cold start problem, the peer response lists were initially seeded with 10 synthetic design arguments generated by an LLM (OpenAI GPT-5). These seeds were generated with prompts designed to mimic the variable quality and structure of typical student writing. Prior work has similarly utilized synthetic peer content in controlled studies to balance ecological validity with experimental control~\cite{peng2025glitter}.

To increase the ecological validity of the corpus as the study progressed, we employed a bootstrapping approach. At the end of each session, participants were asked: \emph{``As mentioned in the informed consent, would you like to anonymously donate your written responses to the study, so that future participants can view them along with the existing ones, attributed to a pseudonym? This won’t affect your compensation.''} All participants consented. These responses were appended to the corpus, meaning later participants encountered a larger and more realistic mix of synthetic and human-authored content.

Importantly, we observed that there were no substantial differences in length or quality between the initial synthetic set ($M=63.5$ words, $M=3.25$ quality score) and the participant-generated set ($M=65.75$ words, $M=3.29$ quality score). Furthermore, because the study used a within-subjects design, any increase in task difficulty due to corpus growth affected both the control and experimental conditions equally for a given participant, preserving the validity of the comparative results.

\subsubsection{Conditions}
Participants used the following two interfaces in counterbalanced order:
\begin{itemize}
  \item \textbf{Interface A (\system{}).} The full \system{} system with Structured Authoring, Structured Exploration with Common-Role Text Segment Highlighting, Peer-Linked Authoring, similarity reordering and search features. All features are enabled throughout the task.
  \item \textbf{Interface B (baseline).} A simplified version of \system{} providing a minimal authoring interface and access to classmates’ responses as a flat list, with only the search feature and the ability to ``Respond'' to a peer's response, similar to a discussion-board platform.
\end{itemize}

For each interface, participants wrote a design argument in response to one of two prior CHI papers (Paper~1, Paper~2). Paper assignment to condition and condition order were both counterbalanced across participants to control for order-related effects such as learning or fatigue and to account for differences in paper difficulty.

\subsection{Procedure}
All studies were conducted remotely via Zoom and facilitated by the first author. Each study took approximately 60 minutes, and each participant received a \$20 Amazon gift card as compensation.

After providing verbal informed consent, participants completed a pre-study survey. They then read the task instructions and received a short primer on design arguments. After that, they accessed the interfaces via a web link (no installation required) and completed two counterbalanced task blocks on the two interfaces. Each task block consisted of a quick walkthrough of the interface, a timed task that lasted 15 minutes with a 5-minute reminder, followed by a short post-condition survey. Participants shared their screens (recorded by the research team) and were encouraged to think aloud during the two tasks. A post-study survey and a 5-minute semi-structured interview concluded the session.

After each task, the design argument written by each participant was recorded and anonymized (attributed only as ``Pn''). These responses were later evaluated by an independent expert (a member of the research team with experience teaching design arguments) who was blinded to the experimental condition. The rater employed a 5-point rubric assessing the clarity and correctness of the identified need and approach, and the coherence and sufficiency of the reasoning connecting the need to the approach. The resulting rating serves as the quality score reported in the next section.

\section{User Study Results}
\subsection{Quantitative Results}
Because our study used a within-subjects design with $N=12$, we analyzed paired responses. 
For each comparison we first tested normality of the within-subject differences (Shapiro--Wilk). 
If normality held, we applied paired $t$-tests and report Cohen’s $d_z$; otherwise, we used Wilcoxon signed-rank tests with rank-biserial $r$. 
To control for multiple comparisons within each family of related items (e.g., six NASA--TLX subscales, five helpfulness items, two post-study items), we applied Holm--Bonferroni corrections to the $p$-values.

\begin{figure}[h]
  \centering
  \includegraphics[width=\linewidth]{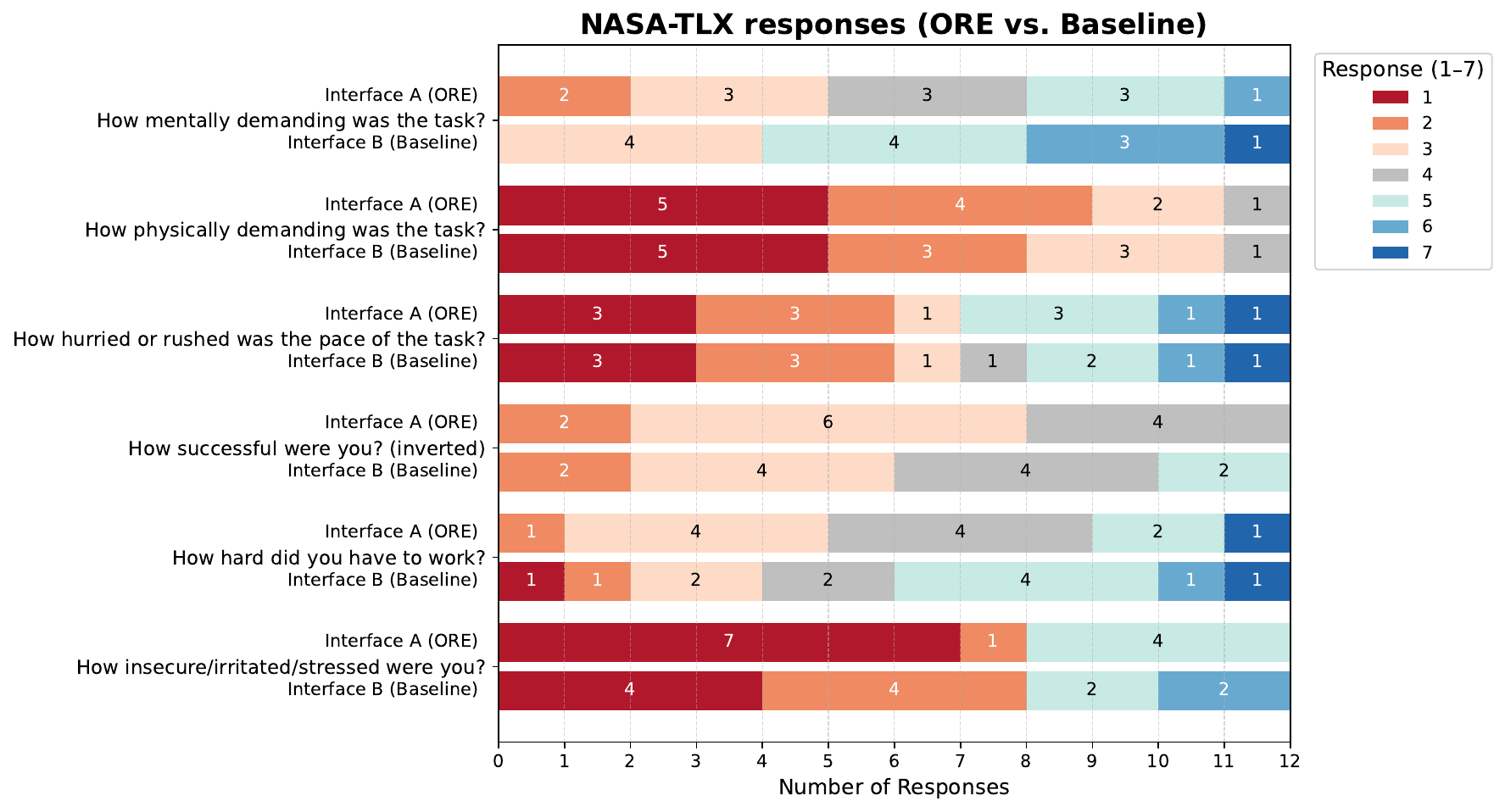}
  \caption{NASA--TLX responses (\system{} vs.\ Baseline). 
  Bars show the distribution of responses (1--7) for each subscale, separated by condition. 
  Numbers inside segments indicate the number of participants (out of 12) who selected that response option. 
  Participants reported significantly lower mental demand with \system{} compared to the Baseline interface. Other subscales and overall workload were not significant after Holm correction.}
  \label{fig:tlx}
\end{figure}

To compare perceived workload between conditions, we analyzed NASA--TLX ratings (1--7 scale; higher indicates greater workload; ``success'' inverted). 
A paired $t$-test showed that participants reported significantly lower \emph{mental demand} with \system{} ($M=3.83, SD=1.27$) than with the Baseline ($M=4.75, SD=1.42$), $t(11)=-3.53, p=.005$, Holm-adjusted $p=.028$, $d_z=1.02$. 
No other subscales were significant after Holm correction, and overall workload did not differ ($p=.213$). 
These results suggest \system{} reduced cognitive demand while leaving overall workload comparable to the baseline (Fig.~\ref{fig:tlx}).

\begin{figure}[h]
  \centering
  \includegraphics[width=\linewidth]{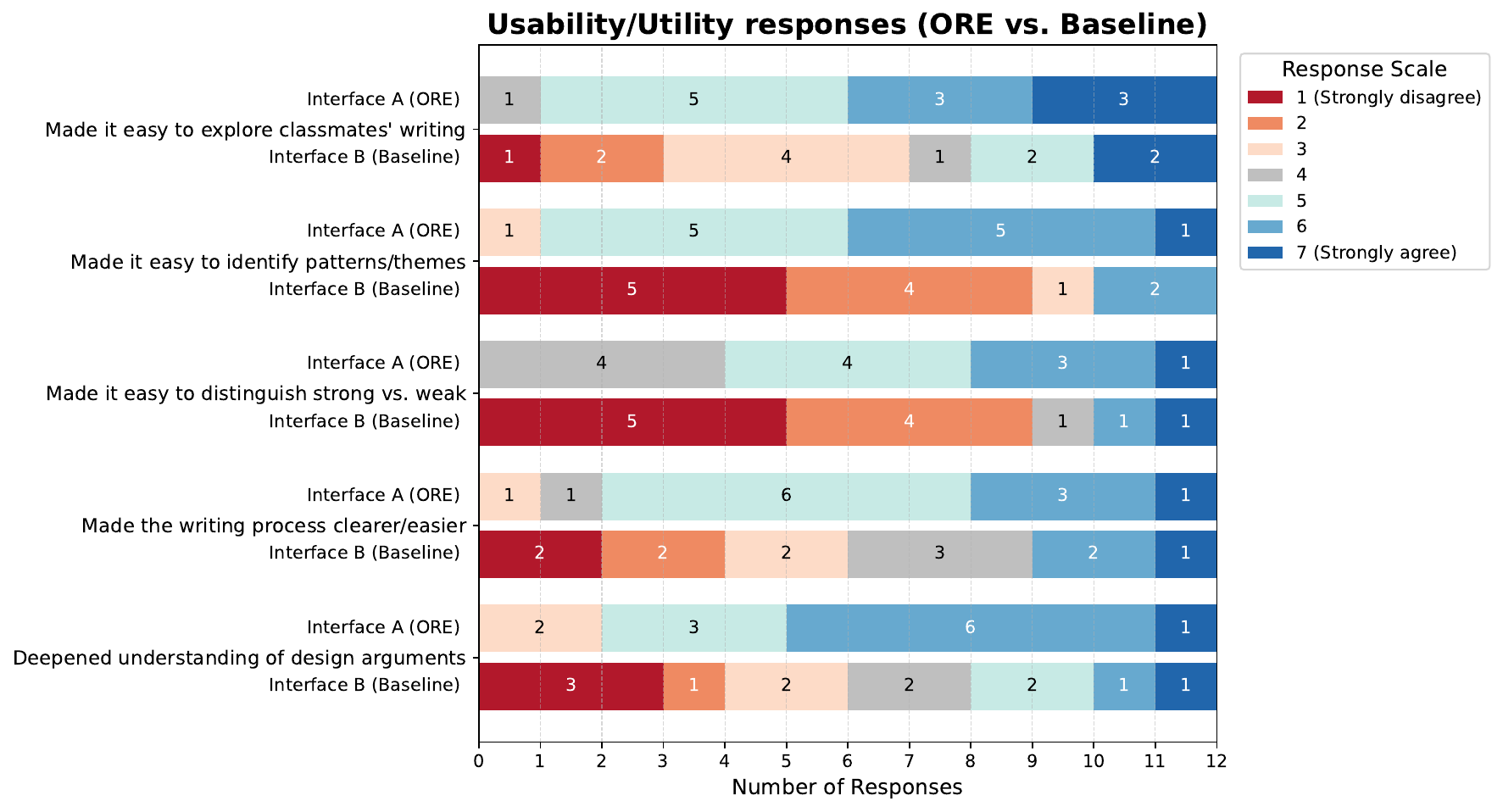}
  \caption{Interface helpfulness ratings (\system{} vs.\ Baseline). 
  Bars show the distribution of responses (1--7) for each question, with counts annotated inside segments. 
  Participants rated \system{} significantly higher on all five items compared to the Baseline. $p$-values remain significant after Holm correction.}
  \label{fig:usability}
\end{figure}

We also compared the interface helpfulness ratings from participants across conditions (1--7 scale). 
Paired $t$-tests (Holm-corrected across five items) showed consistently higher ratings for \system{}: 
\emph{exploring classmates’ writing} ($M=5.67, SD=0.99$ vs.\ $M=3.75, SD=1.91$), $t(11)=3.15$, $p=.009$, Holm $p=.023$, $d_z=0.91$; 
\emph{identifying patterns/themes} ($M=5.42, SD=1.00$ vs.\ $M=2.33, SD=1.83$), $t(11)=4.78$, $p=.001$, Holm $p=.003$, $d_z=1.38$; 
\emph{distinguishing strong vs.\ weak responses} ($M=5.08, SD=1.00$ vs.\ $M=2.50, SD=2.07$), $t(11)=4.76$, $p=.001$, Holm $p=.003$, $d_z=1.37$; 
\emph{making the writing process clearer} ($M=5.17, SD=1.03$ vs.\ $M=3.58, SD=1.98$), $t(11)=2.66$, $p=.022$, Holm $p=.023$, $d_z=0.77$; 
and \emph{deepening understanding of design arguments} ($M=5.33, SD=1.23$ vs.\ $M=3.50, SD=2.02$), $t(11)=3.26$, $p=.008$, Holm $p=.023$, $d_z=0.94$. 
These results indicate that participants consistently perceived \system{} as more helpful and educationally beneficial than the Baseline interface across the writing, peer engagement, and learning stages (Fig.~\ref{fig:usability}).

Additionally, we investigated whether the gradually growing corpus introduced an ordering effect---specifically, whether participants later in the study rated the baseline more harshly due to increased information load. We compared baseline helpfulness ratings between the first half ($N=6$) and second half ($N=6$) of participants. We found no significant difference ($p>0.05$) in any of the ratings, suggesting that the growing corpus size did not systematically skew subjective evaluations of the Baseline interface.

\begin{figure}[h]
  \centering
  \includegraphics[width=\linewidth]{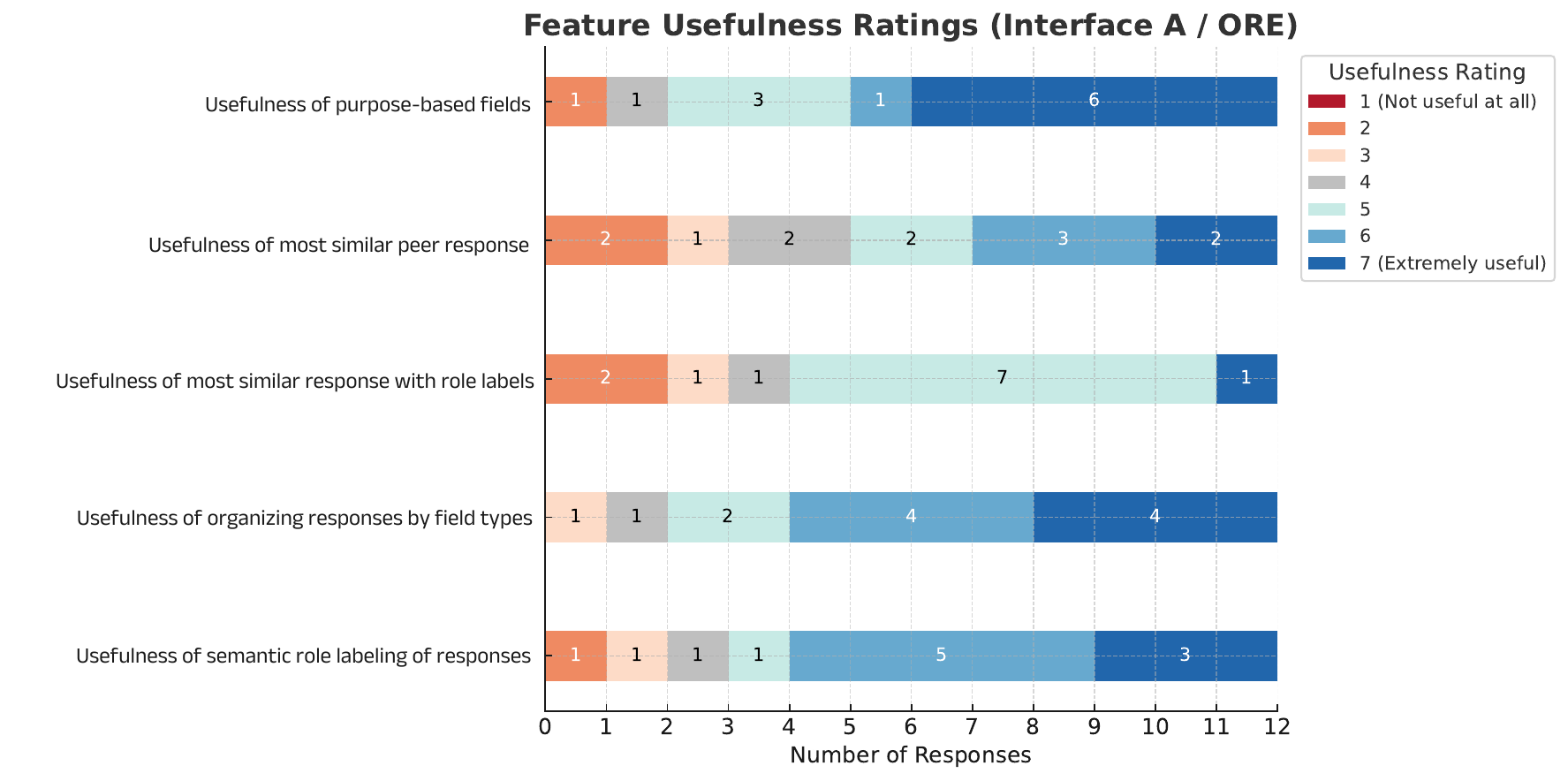}
  \caption{Usefulness ratings of \system{} features (Interface A). 
  Bars show the distribution of ratings (1--7) with counts annotated inside spans. 
  Overall, participants rated all features positively, with purpose-based field organization perceived as the most useful.}
  \label{fig:feature-usefulness}
\end{figure}

Participants also rated the usefulness of \system{} features (1--7 scale). 
Ratings were generally positive: purpose-based fields ($M=5.75, SD=1.60$, median = 6.5) and organizing responses by field types ($M=5.75, SD=1.29$, median = 6.0) received the highest ratings; 
semantic role labeling ($M=5.42, SD=1.62$, median = 6.0) and showing the most similar peer response (with or without role labels; medians = 5.0) received somewhat lower but still favorable evaluations (Fig.~\ref{fig:feature-usefulness}). 

\begin{figure}[t]
  \centering
  \includegraphics[width=\linewidth]{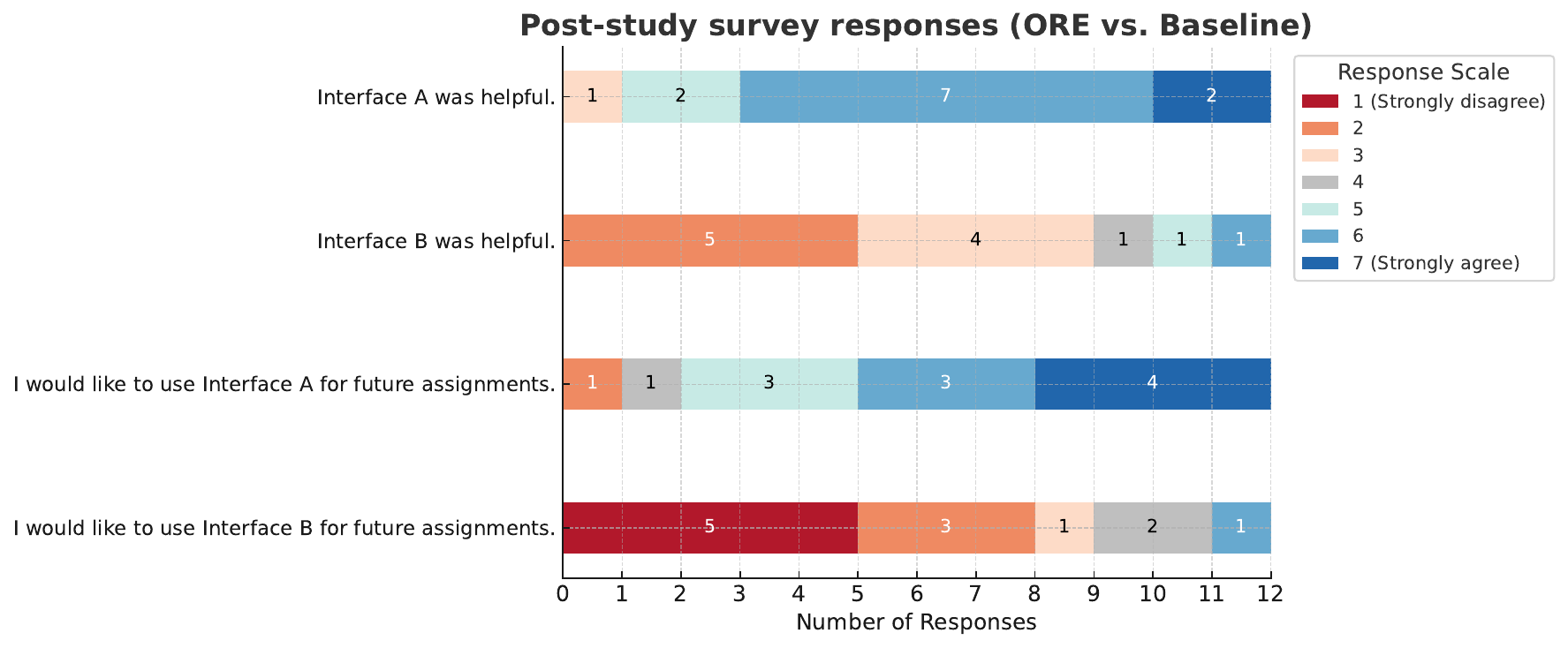}
  \caption{Post–study survey responses. Bars show distributions (1--7) with counts annotated. 
  Participants rated \system{} as significantly more helpful and more desirable for future use than the Baseline.}
  \label{fig:poststudy}
\end{figure}

In the post-study survey, participants rated overall helpfulness and willingness to use the interfaces in future assignments on a 1--7 scale. 
Because the within-subject differences violated normality (Shapiro--Wilk $p{<}.05$), we used Wilcoxon signed-rank tests and report rank-biserial $r$. 
Participants rated \system{} as more helpful ($M=5.75, SD=1.06$) than the Baseline ($M=3.08, SD=1.31$), $W=5.5$, $p=.008$, Holm $p=.016$, $r=.86$, 
and expressed stronger willingness to use \system{} in the future ($M=5.58, SD=1.51$) compared to the Baseline ($M=2.33, SD=1.61$), $W=4.5$, $p=.011$, Holm $p=.016$, $r=.86$. Both difference reached significance after correction. These results indicate strong overall preference for \system{} (Fig.~\ref{fig:poststudy}).

\begin{figure}[t]
  \centering
  \includegraphics[width=\linewidth]{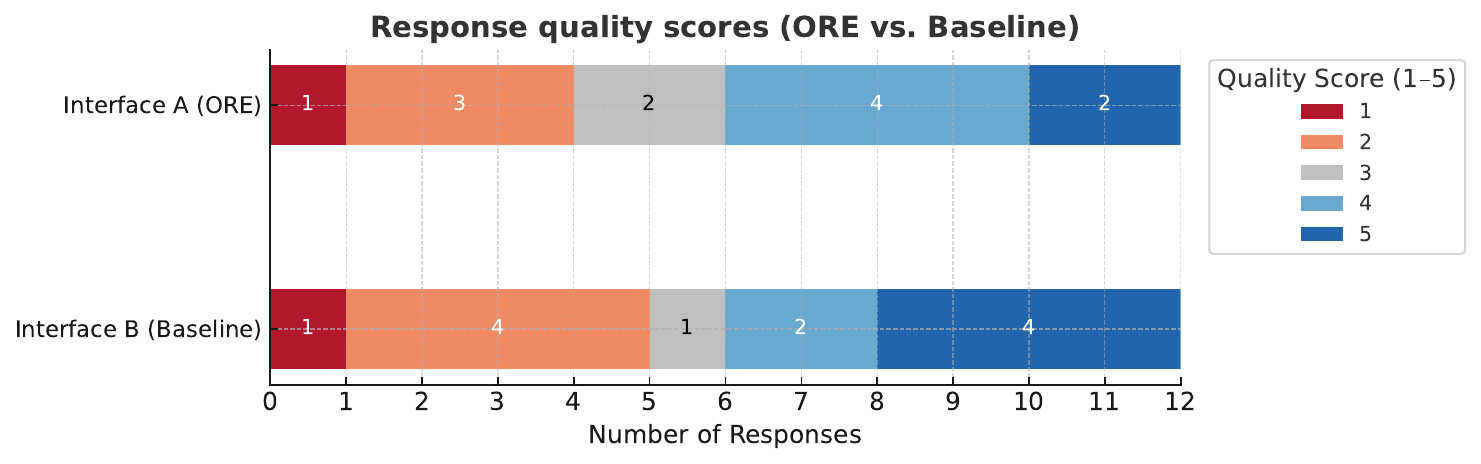}
  \caption{Expert-graded response quality (1--5) for \system{} and Baseline. 
  Bars show distributions with counts annotated. 
  No significant differences were found between the two interfaces.}
  \label{fig:response-quality}
\end{figure}

Finally, we also compared expert-graded response quality scores (1--5 scale) across conditions. 
Responses written with \system{} ($M=3.25, SD=1.29$) and the Baseline ($M=3.33, SD=1.50$) did not differ significantly, $t(11)=-0.15, p=.886, d_z=-0.04$ (Fig.~\ref{fig:response-quality}).

\subsection{Qualitative Results}
We conducted a qualitative analysis of user study transcripts and open-ended survey responses using thematic analysis~\cite{corbin1990basics}. Drawing on participants’ feedback and reflections, the lead researcher iteratively coded the data, revisiting it multiple times to refine labels for consistency and coherence. Using affinity diagramming, high-level themes were then identified and organized. After the thematic structure was finalized, the researcher compiled supporting evidence for each theme and synthesized the results, which the team reviewed to reach agreement on the findings. Below, we group the qualitative results by the key themes that emerged. 

\subsubsection{Structured authoring helped participants organize and articulate ideas}
Many participants described that composing in purpose-oriented fields helped them plan, structure, and refine their writing (P7, P8, P10, P11, P12). 
P7 said, \emph{``it has different cells or [fields] to help me organize my argument, which helps me [write] better answers.''}
P8 considered \emph{``organizing my thoughts in [fields], [and] being able to filter responses in [fields].''} their favorite features. P10 remarked that the interface \emph{``allowed me to break down my writing into sections (which I naturally do) and help promote structure.''}
P11 mentioned, \emph{``I can see this being useful if it were to correspond with the class instructor's rubric. It felt very organized. I also found myself context switching less and skimming more efficiently.''}
P12 added that the [field] roles helped her focus, \emph{``I could focus in on specific parts of the design argument and actually engage with my peers' responses.''}
Together, these accounts indicate that structured authoring reduced cognitive overhead while supporting a clearer writing process.

\subsubsection{Common-role text segment highlighting accelerated sensemaking}
Many praised the color-coded, common-role text segment highlighting and similarity cues for speeding up scanning and comparison (P2, P3, P4, P5, P6, P8, P9, P10, P11, P12). 
P2 said, \emph{``I liked the semantic color-coded highlighting, it made it easier to skim responses and swap out phrases.''}
P3 stated, \emph{``The color-coding in Interface A made it easier for me to read all the responses and get the main structure of the design argument. I could easily navigate the responses and focus on the part where I wanted to get inspiration from.''}
P4 recalled that \emph{``identifying key, underlying themes across students' responses was quite easy''} and that they \emph{``put more thought on the cause-and effect relationships''} as the result of common-role text segment highlighting.
P5 liked \emph{``the option to see similarity via color coding.''} Similarly, P8 observed that they could \emph{``tell a peer is incorrect just by looking at the colors because a complete thesis needs to contain all colors.''} While P6 found the color coding \emph{``overwhelming,''}, they noted that if they were reading \emph{``really big text, then annotations would have been useful.''}
P9 emphasized \emph{``Colors and highlighting''} as the most helpful aspect and said they \emph{``made outlier answers more salient.''} 
P10 noted that \emph{``Semantic labels are extremely helpful; See a lot of people are doing that, maybe I should follow.''}
P12 also discussed how the colors \emph{``made it easier to distinguish strong and weak peer response''} and its potential for facilitating grading.

\subsubsection{Participants asked for persistent context and side-by-side views to reduce context switching}
Several wanted to keep the paper prompt and peer context in view while writing, reducing tab-switching (P3, P4, P8, P10, P12).
P3 proposed, \emph{``a split screen... [so I don’t have to go] back and forth on the tab at the top (between Write and Explore).''}
P4 asked for \emph{``a side panel... to view [more suggestions], instead of only showing the one most relevant after you hit tab/enter.''}
P10 wanted \emph{``to show the question and its template in the editor''} to preserve context.
P12 reflected, \emph{``I would like to have the paper side-by-side... to make sure I wasn't just agreeing with peers without it being true to the paper.''}
These requests point to design opportunities for synthesizing the Write and Explore views to further streamline exploration while composing.

\subsubsection{Accessibility, labeling complexity, and visual semantics could be refined}
Some flagged accessibility and labeling concerns (P1, P9, P11). For example, P1 suggested improving contrast and standards compliance: \emph{``the colors do need to be worked on in terms of WCAG.''}
P9 cautioned that custom color meanings were not always shared, \emph{``different people understand colors differently.''}
P11 found dense labels overwhelming, \emph{``the labeling of some entities was a bit overwhelming. I liked that it could be turned off.''}
These comments highlight the importance of accessible color schemes and progressive disclosure of semantic labels for broader usability.

\subsubsection{Additional AI automation were desired, but  balancing support with autonomy also mattered}
Several suggested lightweight automation for field content (P5, P8, P10, P12) while warning against over-reliance on peers (P12). 
P5 requested \emph{``integration of AI to mark similarity to the paper''} and P12 would like to \emph{``have responses grouped by how similar they are to one another. ''} In addition, P8 and P10 both imagined an additional feature where an LLM provides semantic labels on top of writing \emph{``in real time''} (P8) or \emph{``after pressing a button''} (P10), instead of during exploration only. They noted that it would not only be \emph{``super helpful''} but also make the Write and Explore pages function more consistently. At the same time, P12 reflected on the risk of conformity, \emph{``it could be detrimental if I just agree with my peers' interpretations or AI suggestions. I would like to have the paper side-by-side to cross-check.''} This tension confirms the value of the Instructor-configurable Modes feature, which allows phasing or disabling peer exposure, while also suggesting the need for more customization on the user side to accommodate students' diverse working styles.

\section{Discussion}
\label{sec:discussion}

Overall, \system{} significantly reduced mental demand and was rated significantly higher than the Baseline on all five usability/utility items---making it easier to explore classmates’ writing, identify patterns and themes, distinguish strong from weak responses, make writing clearer/easier, and deepen conceptual understanding. In the post–study survey, participants also judged ORE significantly more helpful and more desirable for future use. Qualitative findings echoed these results: participants described how structured authoring helped them plan and articulate ideas and how common role text segment highlighting accelerated scanning and comparison. Taken together, the results validate the usability and the usefulness of \system{}'s core scaffolds for helping students explore, write, and learn from this open-response assignment.

\subsection{On structured writing and exploration}
Our results suggest students welcome structure, but ``structure'' is task- and data-dependent. For tightly scoped tasks such as paper reviews and design arguments, \emph{purpose-oriented fields} align rhetorical function across writers and enable like-with-like exploration. For divergent tasks such as brainstorming and debates on controversial topics, however, \emph{idea-oriented fields} can foreground distinct framings and alternatives while still allowing comparison by role when appropriate. This dual stance is consistent with process models of writing~\cite{flower1981cognitive,sweller1988cognitive} and with writing-to-learn findings that organization supports understanding \cite{bangertdrowns2004writing}.

Participants favored structured authoring and exploration, yet outcome quality did not improve significantly in a single, short session. One interpretation may be that structure delivered \emph{process gains} (lower mental demand, faster sensemaking, clearer plans) without immediate \emph{product gains} under tight time and limited iteration \cite{bangertdrowns2004writing}.

However, this structural scaffolding introduces a tension: the risk of \emph{over-structuring}. While borrowing structure helps novices internalize genre norms---a key benefit of social learning~\cite{bandura1977social}---it may unintentionally steer students toward surface-level mimicry or homogenization of responses. 
This suggests that the degree of enforced structure should be inversely related to learner expertise: strict, purpose-labeled units scaffold novices by reducing the ``blank page'' problem, while advanced students may require flexible, user-defined structures to encourage divergence. Future designs should therefore make structures \emph{task-tunable} and instructor-configurable, allowing educators to fade structural constraints as students gain proficiency.

\subsection{Beyond common-role text segment highlighting}
Participants valued highlights that align analogous roles across responses. At the same time, some found dense labeling overwhelming, consistent with cognitive theories that warn against excessive signaling \cite{sweller1988cognitive} as well as more recent HCI work that shows constrained highlighting is preferable to unlimited highlighting~\cite{joshi2024constrained}. A practical pattern could be \emph{selective defaulting}: enabling a small, high-utility subset of roles by default, with progressive disclosure to add more roles based on user needs. Additionally, participants also requested \emph{write-time} highlights, which could bring timely, formative feedback on rhetorical coverage at the risk of premature interference or convergence. Staging may reconcile this risk by revealing write-time highlights only after a minimal draft or on explicit request, aligning with productive-failure/engagement principles \cite{kapur2016examining,chi2014icap}. 

\subsection{Implications for Instruction and Adoption}
While our study focused on the student experience, \system{} also introduces new affordances and challenges for instructors. 
Adopting \system{} requires a pedagogical shift from solitary assessment to visible, community-based inquiry, which may encounter friction in contexts where privacy is the default norm. However, this visibility offers distinct advantages for scaling feedback.
By reifying the rhetorical structure of student responses (e.g., separating need thesis from approach thesis), \system{} potentially transforms grading from a linear reading task into a structural verification task. Instructors can visually scan the corpus to identify students who are missing specific rhetorical moves (e.g., an empty or missing ``Evidence'' unit) without reading every word of prose. Furthermore, the ability to view all ``Counter-arguments'' side-by-side enables instructors to rapidly identify common misconceptions or distinct clusters and outliers of student thought, allowing for more targeted class-wide interventions rather than repetitive individual comments.

\subsection{Limitations and future work}
Despite \system{}'s promising contributions, there are several limitations regarding our study design, participant population, and current technical implementation. First, our participants were graduate students who likely possess higher familiarity with peer-informed writing. While this population was appropriate for evaluating the usability of the system's advanced features, future work should investigate how novices, who may struggle more with the cognitive friction of structuring their thoughts, adopt the tool. Relatedly, although \system{} was designed for course use, our evaluation was a lab study intended to validate usability and perceived utility. We plan to conduct a deployment of \system{} in an authentic college-level course to complement lab study findings with in–situ observations of \emph{how} students engage and learn using the system.

Second, our role–aligned highlighting still relies on a human–in–the–loop. Future work could refine the pipeline toward greater automation and robustness, while maintaining labeling reliability and accessibility. Finally, regarding scalability, as the number of purpose and role types scales, simple linear rendering of fields and color-coding may become ambiguous or overwhelming, necessitating hierarchical menus or filtering mechanisms. Similarly, our current approach assumes a degree of structural convergence, but open–response work is diverse and structures are task- and data-dependent. For tasks where student responses are highly divergent or require entirely unique structures, the utility of common-role alignment may diminish. Future iterations could explore the use of adaptive schemas to handle this heterogeneity across disciplines.

\section{Conclusion}
\label{sec:conclusion}

While peer-informed open-response assignments hold the potential to transform classrooms into communities of inquiry, the messiness of unstructured prose often hinders deep social exchange and peer learning at scale. We presented \system{}, a system that addresses this barrier through structured authoring to make rhetorical purposes explicit and comparable by design, and structured peer response exploration with role-based highlighting to facilitate sensemaking at scale. Our evaluation confirms that these scaffolds significantly reduce the mental demand of writing and sensemaking, helping students better identify patterns and themes in each other's writing. Ultimately, this work opens the door for new classes of structured, purpose-centered writing and sensemaking interfaces that can unlock the full pedagogical potential of collaborative knowledge building and social learning.

\bibliographystyle{ACM-Reference-Format}
\bibliography{main}

\appendix

\section{User Study Task}
\label{sec:task}
Task Instructions:

In this study, you will complete two short open-ended writing tasks on two different interfaces. 

Imagine you're completing a course assignment where you’re asked to write a design argument in response to a research paper. A few of your ``classmates'' have already submitted their responses, and you can view them as examples. Your classmates are learning too—they may or may not fully understand design arguments yet either. 

Your task is to 1) read the Abstract (and optionally the Introduction section) of the research paper, 2) review your classmates’ responses, and 3) then write the best short design argument (2-5 sentences) you can.

What is a Design Argument?

A design argument is a short essay that makes the case for a proposed tool or approach by articulating the need for the tool and how the proposed approach uniquely addresses the need in context. A more detailed description can be found here.

You will have 15 minutes to complete each task, including both exploring others’ responses and writing your own response. 

Remember: 

- There is no right or wrong answer.

- The tasks are a test of the system’s effectiveness, not your intelligence.

- It’s okay to use the language of the authors (i.e. copy-and-paste from the original paper).

- You can disagree with the author and/or your classmates!

\section{Additional Screenshot}
\begin{figure}
    \centering
    \includegraphics[width=\linewidth]{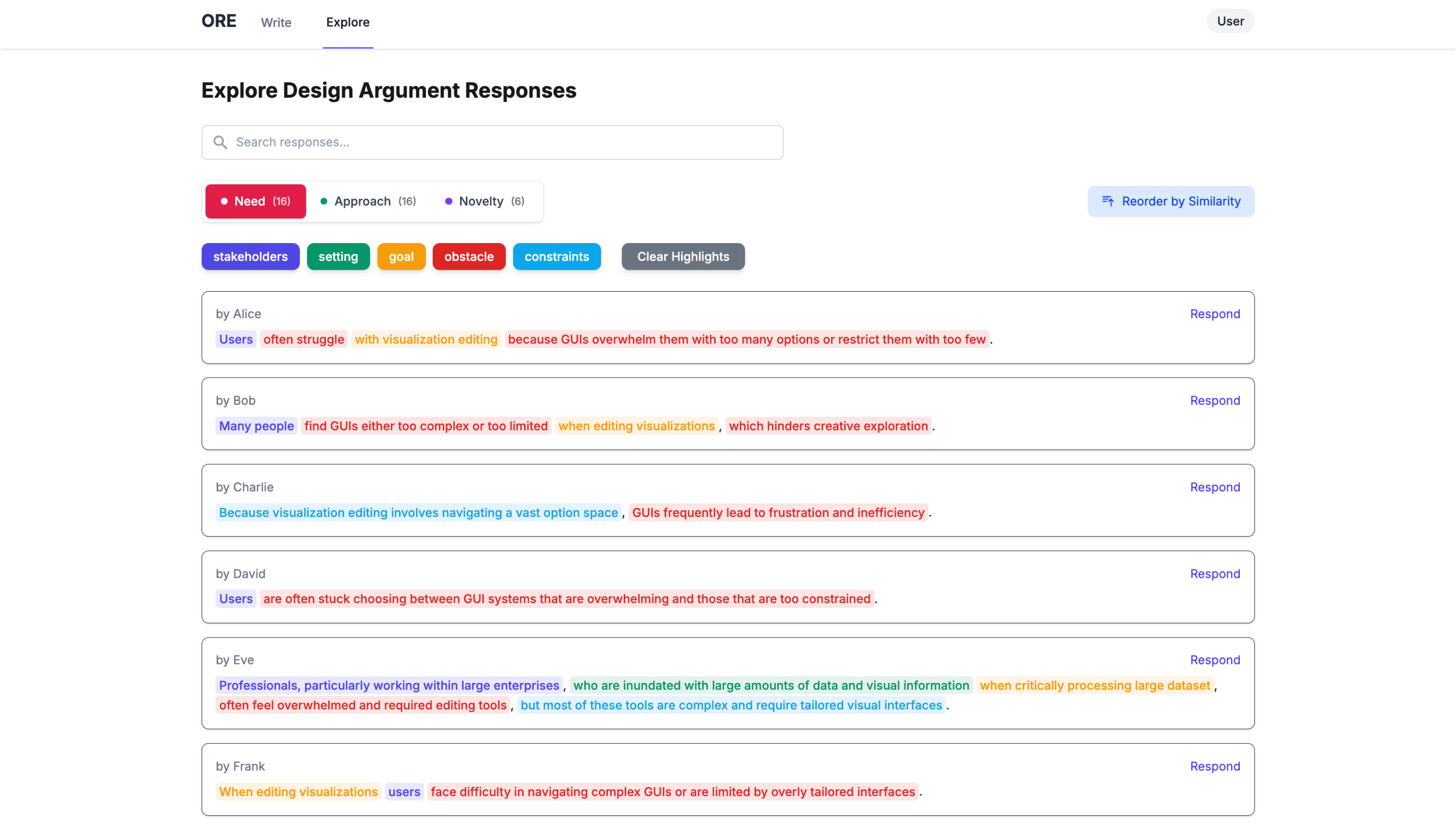}
    \Description[]{}
    \caption{\textbf{Explore} view with ``Need'' field type selected}
    \label{fig:explore2}
\end{figure}

\end{document}